\documentclass[10pt]{article}
\usepackage{graphicx}
\usepackage{amsmath,amssymb,amsthm,amsfonts}
\usepackage{amssymb}
\newtheorem{thm}{Theorem}

\newtheorem{lem}{Lemma}

\theoremstyle{definition}

\theoremstyle{remark}

\numberwithin{equation}{section}

\DeclareMathSymbol{\C}{\mathalpha}{AMSb}{"43}

\newcommand{\eps}{\varepsilon}

\newcommand{\R}{{\mathbb{R}}}
\newcommand{\h}{{\mathcal{H}}}
\newcommand{\inte}{\int_{\mathbb{R}^2}}

\newcommand{\bsub}{\begin{subequations}}
\newcommand{\esub}{\end{subequations}$\!$}

\begin{document}

\title{On the Mass Concentration for Bose-Einstein Condensates
  with Attractive Interactions}

\author{Yujin Guo\thanks{Wuhan Institute of Physics and Mathematics,
    Chinese Academy of Sciences, P.O. Box 71010, Wuhan 430071,
    P. R. China.  Email: \texttt{yjguo@wipm.ac.cn}. Y. Guo is partially supported by  NSFC grants 11241003 and 11322104.}
\, and\, Robert Seiringer\thanks{IST Austria, Am Campus 1, 3400 Klosterneuburg, Austria.  Email: \texttt{robert.seiringer@ist.ac.at}. R. Seiringer
    is partially supported by the Natural Science and Engineering
    Research Council of Canada.}
}

\date{\today}

\smallbreak \maketitle

\begin{abstract}
  We consider two-dimensional Bose-Einstein condensates with
  attractive interaction, described by the Gross-Pitaevskii
  functional. Minimizers of this functional exist only if the
  interaction strength $a$ satisfies $a < a^*= \|Q\|_2^2$, where $Q$
  is the unique positive radial solution of $\Delta u-u+u^3=0$ in
  $\R^2$. We present a detailed analysis of the behavior of
  minimizers as $a$ approaches $a^*$, where all the mass concentrates
  at a global minimum of the trapping potential.
\end{abstract}

\vskip 0.1truein

\noindent {\it Dedicated to Nassif Ghoussoub on the occasion of his $60^{\rm th}$ birthday.}

\vskip 0.2truein


\vskip 0.2truein


\vskip 0.2truein

\section{Introduction and Main Results}

The phenomenon of Bose-Einstein condensation (BEC) has been
investigated intensively since its first realization in cold atomic
gases \cite{Anderson,Ketterle}. In these experiments, a large number of (bosonic) atoms are
confined to a trap and cooled to very low temperature. Condensation of
a large fraction of particles into the same one-particle state is
observed below a critical temperature.

These Bose-Einstein condensates display various interesting quantum
phenomena \cite{D,Cooper,Fetter,Zwerger}, such as the appearance of
quantized vortices in rotating traps, the effective lower dimensional
behavior in strongly elongated traps, etc.  The forces between the
atoms in the condensates can be either attractive or repulsive. In the
attractive case, the system collapses if the particle number increases
beyond a critical value; see, e.g., \cite{Hulet1,Hulet2,HM,KM,Hulet3} or
\cite[Sec.~III.B]{D}.  Our main interest in the present paper is to
investigate the details of this collapse.

We study Bose-Einstein condensates with attractive
interactions in two dimensions, described by the Gross-Pitaevskii (GP) energy functional \cite{G60,G63,P}.
In suitable units, the GP functional  is given by
\begin{equation}
  E_a(u):=\int_{\R ^2} \big(|\nabla
  u(x)|^2+V(x)|u(x)|^2\big)dx-\frac{a}{2}\int_{\R ^2}|u(x)|^4dx\, , \quad  u\in \h \,, \label{f}
\end{equation}
 where $ a>0$ describes the strength of the attractive interactions,
 and $\h$ is defined as
\begin{equation}
   \h := \Big \{u\in  H^1(\R ^2):\ \int _{\R ^2}  V(x)|u(x)|^2 dx<\infty \Big\}\,   \label{H}
\end{equation}
with associated norm $\|u\|_\h=\{\inte \big(|\nabla u|^2+[1+V(x)]|u(x)|^2 \big)dx\}^{\frac{1}{2}}$.
We note that in the case of repulsive interactions (corresponding to
$a<0$ in (\ref{f})), the GP functional can be rigorously derived from
the quantum many-body problem in a suitable low-density limit \cite{LSS,LSY,LSY2d,LS}. Its
validity in the attractive case remains an open problem, however.

We are interested in minimizers of (\ref{f}) under the unit mass constraint
\begin{equation}\label{norm}
\int_{\mathbb{R} ^2} |u(x)|^2dx=1 \,.
\end{equation}
Alternatively, one may want to impose the constraint $\int_{\mathbb{R}
  ^2} |u(x)|^2dx=N$, with $N$ the particle number, but this latter
case can easily be reduced to the previous one, by minimizing under the constraint (\ref{norm}) but simply replacing $a$ by $Na$.  Hence we prefer to work with (\ref{norm}) instead.

We assume that the function $V:\R^2\to \R$ is locally bounded and
satisfies $V(x)\to\infty$ as $|x|\to\infty$. By adding a suitable
constant, we may impose the condition $\inf_{x\in \R^2} V(x) = 0$
without loss of generality.
We define the GP energy to be
\begin{equation}\label{def:ea}
e(a):=\inf _{\{u\in \h, \, \|u\|^2_2=1 \} } E_a(u) \, .
\end{equation}
Note that, without loss of generality, we can restrict the minimization to non-negative functions, since $E_a(u)\geq E_a(|u|)$ for any $u\in \mathcal{H}$. This follows from the fact that $|\nabla |u||\leq |\nabla u|$ a.e. in $\R^2$.

We start by investigating the finiteness of $e(a)$. As an infimum over affine-linear functions, $e(a)$ is concave, and it is clearly also decreasing in $a$. 
Using the fact that $\int |\nabla u|^2$ and $\int |u|^4$ behave the same under ($L^2$-preserving) scaling of $u$, it is easy to see that either $e(a)\geq 0$ or $e(a) =
-\infty$. Hence a simple variational argument yields the existence of
an $a^*\geq 0$ such that $e(a) = -\infty$ for $a> a^*$. It turns out that the value of
$a^*$ can be determined by solving the nonlinear scalar field
equation
\begin{equation}
-\Delta u+ u-u^3=0\  \mbox{  in } \  \R^2,\  \mbox{ where }\  u\in H^1(\R ^2).  \label{Kwong}
\end{equation}
It is well-known \cite{GNN,Li,mcleod,K} that, up to translations,
(\ref{Kwong}) admits a unique positive solution, which can be taken to
be radially symmetric about the origin. We shall denote it by
$Q$. Moreover, we recall from \cite{W} the following
Gagliardo-Nirenberg inequality
\begin{equation}\label{GNineq}
\inte |u(x)|^4 dx\le \frac 2 {\|Q\|_2^{2}} \inte |\nabla u(x) |^2dx \inte |u(x)|^2dx ,\   \  u \in H^1(\R ^2)\,,
\end{equation}
where equality is achieved for $u(x) = Q(x)$. This inequality can be used to obtain the following theorem concerning the existence and non-existence of minimizers for the minimization problem (\ref{def:ea}).

\begin{thm}\label{thm1} Let $Q$ be the unique positive radial solution of (\ref{Kwong}). Suppose $V\in L^\infty_{\rm loc}(\R^2)$ satisfies $\lim_{|x|\to\infty} V(x) = \infty$ and $\inf_{x\in \R^2} V(x) =0$. Then
\begin{enumerate}
\item If $0\leq a< a^*:=\|Q\|^2_2$, there exists at least one minimizer for (\ref{def:ea}).
\item If  $a \ge a^*:=\|Q\|^2_2$, there is no minimizer for (\ref{def:ea}).
\end{enumerate}
Moreover, $e(a) > 0$ for $a<a^*$, $\lim_{a\to a^*} e(a) = e(a^*) = 0$, and $e(a) = -\infty$ for $a>a^*$.
\end{thm}

The proof of the existence of minimizers in the case $0\le a<a^*$ follows standard arguments (see \cite{R,Z}) and we include it here for completeness. A numerical computation \cite{W} yields  $\|Q\|^2_2=2\pi \times 1.86225 \cdots$. It is not difficult to obtain analytical bounds as well. We shall demonstrate in Lemma~\ref{Kwong:lem1} below  that
\begin{equation}
 2\pi\le \|Q\|^2_2 \le  2\pi e \ln 2 \approx 2\pi \times 1.88417\cdots \ .  \label{estimate:a^*}
\end{equation}

Note that since the parameter $a$ in (\ref{f}) has to be interpreted as
particle number times interaction strength, as discussed after
Eq.~(\ref{norm}) above, the existence of the threshold $a^*$ described
in Theorem \ref{thm1} yields the existence of a critical particle number for
collapse of the Bose-Einstein condensate \cite{D}. Theorem \ref{thm1} also implies that the trap shape does not affect the critical particle number (compare with \cite{Y}).

If $u$ is a non-negative minimizer of (\ref{f}) for $a<a^*$, it satisfies the GP equation
\begin{equation}
-\Delta u(x) +V(x)u(x) =\mu u(x) +au(x)^3\quad \mbox{  in } \quad \R^2  \label{eqn}
\end{equation}
for a suitable Lagrange parameter $\mu$. From this equation one can
deduce exponential decay of $u$ via standard techniques. Moreover, the
maximum principle implies that $u$ is strictly positive, and elliptic
regularity yields smoothness properties of $u$ depending on the
smoothness assumptions on  $V$. These results are standard (see, e.g., \cite{C})
and we shall not investigate the details here.

\bigskip

{\em Our main result} concerns the behavior of minimizers $u_a$ of
(\ref{f}) as $a$ approaches the critical value $a^*$ from below. Since
$e(a^*)=0$, it is easy to see that $\int_{\R^2} V(x) |u_a(x)|^2 dx \to
0 = \inf_{x\in \R^2} V(x)$ as $a\to a^*$, hence this behavior depends
on the behavior of $V$ near its minima. The functions $u_a$ can be
expected to concentrate at the flattest minimum of $V$. If $V$ has a
unique minimum, $|u_a(x)|^2$ converges to a $\delta$-function located
at this minimum.

In the following, we shall assume that the trap potential $V$ has
$n\geq 1$ isolated minima, and that in their vicinity $V$ behaves like
a power of the distance from these points. More precisely, we shall
assume that there exist $n\geq 1$ distinct points $x_i\in \R^2$ with
$V(x_i)=0$, while $V(x)>0$ otherwise. Moreover, there are numbers
$p_i>0$ and a constant $C>0$ such that
\begin{equation}\label{as:v}
V(x) =  h(x) \prod_{i=1}^n |x-x_i|^{p_i} \quad \text{with $C < h(x) < 1/C$ for all $x\in \R^2$.}
\end{equation}
We also need to assume that $\lim_{x\to x_i} h(x)$ exists for all
$1\leq i \leq n$. 

Our method certainly allows to relax these
assumptions in various ways. For instance, a very rapid increase at infinity (which could lead to $Q\not\in \mathcal{H}$) could be handled by introducing suitable additional cut-off functions. 
For the sake of simplicity we shall
not strive to cover the most general class of trap potentials here. We note that new ingredients in the proof are needed to treat the case of radially symmetric, ring-shaped potentials (with a continuum of minima) and the corresponding analysis will be presented elsewhere \cite{GZZ}.

Let $p = \max \big\{p_1,\dots ,p_n\big\}$, and let $\lambda_i \in (0,\infty]$ be given by
\begin{equation}\label{def:li}
  \lambda_i = \left( \frac p 2  \int_{\R^2} |x|^p Q(x)^2 dx \,  \lim_{x\to x_i} \frac{V(x)}{|x-x_i|^p} \right)^{\frac 1{2+p}} \,.
\end{equation}
Define $\lambda = \min\big\{\lambda_1,\cdots ,\lambda_n\big\}$ and let
\begin{equation}\label{def:Z}
\mathcal{Z}:=\big\{x_i:\, \lambda_i=\lambda\big\}
\end{equation}
denote the locations of the flattest global minima of $V(x)$.

\begin{thm}\label{thm3}
  Suppose $V$ satisfies the assumptions above, and let $u_a$ be a
  non-negative minimizer of (\ref{f}) for $a<a^*$. Given a
  sequence $\{a_k\}$ with $a_k\nearrow a^*$ as $k\to\infty$, there
  exists a subsequence (still denoted by $\{a_k\}$) and an $x_0 \in \mathcal{Z}$ such that
\begin{equation}
\lim_{k\to \infty} (a^*-a_k)^{\frac{1}{2+p}} u_{a_k}\left(x_0 + x  (a^*-a_k)^{\frac{1}{2+p}}\right) =  \frac { \lambda Q(\lambda x)}{\|Q\|_2}  \label{con:a}
\end{equation}
strongly in $L^q(\R^2)$ for $2\leq q<\infty$.
\end{thm}

The theorem gives a detailed description of the behavior of GP minimizers close to the critical coupling strength.
As $a\to a^*$, a minimizer $u_a$ of (\ref{f}) behaves like
\begin{equation}
 u_{a}(x) \approx \frac{\lambda }{\|Q\|_2(a^*-a_k)^\frac{1}{2+p}}Q\Big(\frac{ \lambda (x-x_0)}{(a^*-a_k)^\frac{1}{2+p}}\Big)\,,
\end{equation}
with $x_0$ a minimum of $V(x)$, and $\lambda$ the smallest of the
values $\lambda_i$ defined in (\ref{def:li}). Such an equality can, in
general, hold only for a subsequence. If $x_0$ is unique, however,
i.e., $|\mathcal{Z}|=1$, it is not necessary to go to a subsequence,
and the convergence (\ref{con:a}) holds for any sequence.

If the trap potential $V$ has a symmetry, e.g.,  $V(x)=\prod
_{i=1}^n|x-x_i|^p$ with $p>0$ and the $x_i$ arranged on the vertices
of a regular polygon, Theorem~\ref{thm3} establishes the {\em symmetry
  breaking} occurring in the GP minimizers. There exists an $a_*$,
with $0< a_* <a^*$, such that for $a_* < a < a^*$, the GP functional (\ref{f}) has
(at least) $n$ different non-negative minimizers, each of which
concentrates at a specific global minimum point $x_i$. 

One can show that symmetry breaking can only occur for $a$
sufficiently large. That is, for small enough $a$, there is always a
unique minimizer (up to multiplication by a constant phase), as in the
case $a=0$. This can be proved using the technique employed in the
proof of Theorem~2 in \cite{Schnee} (see also \cite{R} and Theorem~1.1 in
\cite{M}). We note that the  symmetry breaking bifurcation for ground states for nonlinear Schr\"odinger/Gross-Pitaevskii equations has been studied in detail in the literature, see, e.g., \cite{J,K08,K11}. 

Our proof of Theorem~\ref{thm3} is based on precise estimates on the GP energy $e(a)$.
In fact, we shall show that
\begin{equation}
e(a) \approx   (a^*- a)^{p/(2+p)}  \frac{\lambda^2}{a^*} \frac{p+2}{p}  \quad \text{as $a\to a^*$}\,.
\end{equation}
Note that the convergence in (\ref{con:a}) also implies that
\begin{equation}
\inte |u_a(x)|^4 dx \approx  2  (a^*- a)^{-2/(2+p)} \frac{\lambda^2}{a^*} \quad \text{as $a\to a^*$}
\end{equation}
for a minimizer $u_a$.

The results in the present paper can be extended not
only to more general trapping potentials $V$, but also to space
dimensions $d$ different from $2$, if the exponent $4$ in the last
term in (\ref{f}) is replaced by $p= 1+\frac{4}{d}$. Previous results
in \cite{M} were restricted to the subcritical case $p< 1+\frac 4 d$,
where one studies the behavior of minimizers as $a\to \infty$. The
case of a non-local nonlinearity was considered in \cite{Schnee}.
Concentration phenomena have also been studied elsewhere in different
context. For instance, there is a considerable literature on
concentration phenomena of positive ground states of the elliptic
equation
\begin{equation}
h^2\Delta u(x)-V(x)u(x)+ u(x)^p=0
\quad \mbox{in }\ \R^d \label{4con:a}
\end{equation}
as $h\to 0^+$, see \cite{floer,oh,Wang,BWang} and references therein.

In the remainder of this paper, we shall give the proof of Theorems~\ref{thm1} and~\ref{thm3}.

\section{Proof of Theorem~\ref{thm1}}

In this section, we prove Theorem \ref{thm1} on the threshold $a^*$, which can be defined as
\begin{equation}
a^*=\sup\big\{a >0\ |\ (\ref{def:ea}) \ {\rm possesses \ at \ least\
one\ minimizer} \big\} \,.
\end{equation}
Recall that $Q$ is the unique positive radial solution of (\ref{Kwong}), and that
 $\frac{1}{2}\int _{\R^2}Q(x)^2dx$ is the minimum of the energy functional
\begin{equation}
I(u)= \frac{\inte |\nabla u(x) |^2dx \inte |u(x)|^2dx}{\inte |u(x)|^4dx},\  \mbox{ where }\  u\in H^1(\R ^2).  \label{Kwong:2}
\end{equation}
The function $Q$ is decreasing away from the origin, and \cite[Prop.~4.1]{GNN}
 \begin{equation}
Q(x) \, , \ |\nabla Q(x)| = O(|x|^{-\frac{1}{2}}e^{-|x|}) \quad \text{as  $|x|\to \infty$.}  \label{4:exp}
\end{equation}

We first discuss the following bounds on $\|Q\|^2_2$.

\begin{lem}\label{Kwong:lem1}  Let $Q$ be the unique positive radial solution of (\ref{Kwong}). Then we have
\begin{equation}
2\pi\le\int _{\R^2}Q(x)^2dx\le    2\pi e \ln 2 \approx 2\pi \times 1.88417\cdots .  \label{Kwong:E}
\end{equation}
\end{lem}

\noindent{\bf Proof.} The lower bound was proved in~\cite{L}.  To
obtain the upper bound, we consider the trial functions $u_\gamma(x) =
e^{-|x|^\gamma/2}$ for $\gamma>0$. Simple calculations yield
$$
\|u_\gamma\|_2^2  = \pi \Gamma(1+2/\gamma) =  2^{2/\gamma} \|u_\gamma\|_4^4 \ , \quad \|\nabla u_\gamma\|_2^2 = \frac{\pi\gamma}2\,,
$$
and hence
$$
\int_{\R^2} Q(x)^2 dx \leq 2 \inf_{\gamma>0} I(u_\gamma) = \inf_{\gamma>0} \pi \gamma 2^{2/\gamma} = 2\pi e \ln 2\,,
$$
where the infimum is attained at $\gamma=\ln 4$.
\qed
\bigskip

The accuracy of our upper bound in Lemma \ref{Kwong:lem1} can be
observed from the numerical result $\|Q\|^2_2=2\pi \times 1.86225
\cdots$ in \cite{W}.

The following compactness result is well known, see, e.g., \cite{Adams}.

\begin{lem}\label{2:lem1}  Suppose $V \in L_{\rm loc}^\infty(\R^2)$ with $\lim_{|x|\to \infty} V(x) = \infty$.  If  $2\le q<\infty$, then the embedding $\h \hookrightarrow L^{q}(\R^2)$ is compact.
\end{lem}

This compactness property, together with the Gagliardo-Nirenberg inequality (\ref{GNineq}), allows us to prove Theorem~\ref{thm1}. We shall first show that (\ref{def:ea})  admits at least one minimizer provided that $0\le a<\|Q\|^2_2$.
The proof of this fact is essentially the same as the one in \cite{Z}, where the special case $V(x)=|x|^2$ was considered (see also \cite{R}). We give it here for completeness.

If $u\in\h $ and $\|u\|^2_2=1$, then for all $0\leq a< \|Q\|^2_2$ we observe from (\ref{GNineq}) and the positivity of $V$ that
\begin{align}\nonumber
 E_a(u) & \ge \Big(1-\frac{a}{\|Q\|^2_2}\Big)\inte |\nabla u|^2dx + \inte V(x)|u(x)|^2 dx \\
&\ge \Big(1-\frac{a}{\|Q\|^2_2}\Big)\inte |\nabla u|^2dx \,,
\label{key:min}
\end{align}
which implies that $E_a(u)$ is bounded from below. Let $\{u_n\}\in \h$ be a sequence satisfying $\|u_n\|_2=1$ and $\lim_{n\to \infty} E_a(u_n) = e(a)$. Because of (\ref{key:min}), we see that both
 $\inte |\nabla u_n(x)|^2dx$ and $\inte V(x)|u_n(x)|^2dx$ are uniformly bounded in $n$.  By the compactness of Lemma \ref{2:lem1}, we can extract a subsequence such that
\[
u_n\rightharpoonup u\ \text{weakly in $\mathcal{H}$}, \quad  u_n\to u\ \text{strongly in $L^q(\R^2)$, \ $2\le q<\infty$}
\]
for some $u\in \mathcal{H}$. We conclude that $\inte |u(x)|^2dx=1$ and $E_a(u)=e(a)$, by weak lower semicontinuity. This implies the existence of  minimizers for any $0\leq a<\|Q\|^2_2$.


To prove that there is no  minimizer for (\ref{def:ea}) as soon as $a \ge \|Q\|_2^2$, we proceed as follows. Choose a non-negative $\varphi \in C_0^\infty(\R^2)$ such that $\varphi(x) = 1$ for $|x|\leq 1$. For $x_0\in \R^2$, $\tau>0$ and $R>0$, let 
\begin{equation}\label{def:trial}
u(x) =  A_{R,\tau} \frac{\tau}{\|Q\|_2} \varphi((x-x_0)/R) Q(\tau (x-x_0)) \,,
\end{equation}
where $A_{R,\tau}>0$ is chosen so that $\inte u(x)^2dx=1$. By scaling, $A_{R,\tau}$ depends only on the product $R\tau$, and we have $\lim_{\tau\to \infty} A_{R,\tau} = 1$.  In fact,
\begin{equation}\label{up1}
\frac 1{A_{R,\tau}^{2}} = \frac{1}{\|Q\|_2^2} \int_{\R^2} Q(x)^2 \varphi(x/(\tau R))^2 dx = 1 + O( (R\tau)^{-\infty}) \quad\text{as $R\tau \to \infty$}
\end{equation}
because of the exponential decay of $Q$ in Eq.~(\ref{4:exp}). Here we use the notation $f(t) = O(t^{-\infty})$ for a function $f$ satisfying $\lim_{t\to \infty} |f(t)| t^s = 0$ for all $s>0$.
In the following, we could set $R=1$, for instance.

Using the exponential decay of both $Q$ and $\nabla Q$, we also have 
\begin{align}\nonumber
&\inte |\nabla u(x) |^2dx- \displaystyle\frac{a}{2}\inte u(x)^4dx\\ 
& =  \frac{\tau^2}{\|Q\|_2^2} \left[\inte |\nabla Q(x)|^2dx- \displaystyle\frac{a}{2\|Q\|_2^2}\inte Q(x)^4dx + O((R\tau)^{-\infty} ) \right] \quad \mbox{as}\quad R\tau\to\infty . \label{fh}
\end{align}
Since $\|\nabla Q\|_2^2 = \frac 12 \|Q\|_4^4$, we further have  
 \begin{equation}
(\ref{fh})  = \displaystyle\frac{\tau^2}{2\|Q\|_2^2} \left[\left( 1-\frac{a}{\|Q\|_2^2}\right) \inte Q(x)^4dx + O( (R\tau)^{-\infty}) \right]\quad \mbox{as}\quad R\tau\to\infty .
 \label{W:est1}
\end{equation}
On the other hand, since the function $x \mapsto V(x) \varphi((x-x_0)/R)^2$ is bounded and has compact support, the convergence 
\begin{equation}\label{W:est2}
\lim_{\tau \to \infty} \int_{R^2} V(x) u(x)^2 dx = V(x_0) 
\end{equation}
holds for almost every $x_0\in \R^2$ \cite{Lieb}.  

For $a >\|Q\|_2^2$, it follows from (\ref{W:est1}) and (\ref{W:est2}) that
$$
e(a) \leq \lim_{\tau \to \infty} E_a(u) = - \infty\,.
$$
This implies that  for any $a >\|Q\|_2^2$, $e(a)$   is unbounded from below, and the non-existence of  minimizers is therefore proved. In the case $a =\|Q\|_2^2$, (\ref{W:est1}) and (\ref{W:est2}) show in combination that $e(a) \leq V(x_0)$. This holds for almost every $x_0$; taking the infimum over $x_0$ yields $e(a)\leq 0$. There is, in fact, equality in this case, as (\ref{key:min}) shows. Suppose now that there exists a minimizer $u$ at $a=\|Q\|_2^2$. As pointed out in the Introduction, we can assume $u$ to be non-negative. We would then have 
$$
\int_{\R^2} V(x) |u(x)|^2 dx = \inf_{x\in \R^2} V(x) = 0
$$
and 
$$
\int_{\R^2} |\nabla u(x)|^2 = \frac 12 \int_{\R^2} |u(x)|^4 dx\,.
$$
This is a contradiction, since for the first equality $u$ would have to have compact support, while for the second one it has to be equal to (a translation of) $Q$. 

This completes the proof of the first part of Theorem \ref{thm1}. To
prove the stated properties of the GP energy $e(a)$, note that
(\ref{key:min}) implies that $e(a)>0$ for $a<a^*=\|Q\|_2^2$. We have already
shown that $e(a^*)=0$ and $e(a) = -\infty$ for $a>a^*$, hence it remains to show that $\lim_{a\to a^*} e(a) = 0$. This follows easily from (\ref{W:est1}) and (\ref{W:est2}), by first taking $a\to a^*$, followed by $\tau\to \infty$. This implies that $\limsup_{a\to a^*} e(a) \leq V(x_0)$ which, after taking the infimum over $x_0$, yields the result.
\qed


\section{Proof of Theorem~\ref{thm3}}

We shall now restrict our attention to trap potentials $V$ satisfying (\ref{as:v}).
We have already shown in Theorem~\ref{thm1} that $e(a)\searrow 0$ as $a\nearrow a^*$, where $e(a)$ is the GP energy defined in (\ref{def:ea}). In the following we shall derive refined estimates on $e(a)$.

\begin{lem}\label{4:lem1}
Suppose $V$ satisfies (\ref{as:v}). Then there exist two positive constants $m<M$, independent of $a$, such that
\begin{equation}
m(a^*-a)^{\frac{p}{p+2}}\le e(a)\le M(a^*-a)^{\frac{p}{p+2}}\ \ \mbox{for}\ \ 0\leq a\leq  a^*,
\label{4:con:1}
\end{equation}
where $p>0$ is defined before~(\ref{def:li}).
\end{lem}

\noindent{\bf Proof.} Since $e(a)$ is decreasing and uniformly bounded for $0\leq a\leq a^*$, it suffices to consider the case when $a$ is close to $a^*$. We start with the lower bound.
From (\ref{GNineq}) we infer that, for any $\gamma >0$ and $u\in \mathcal{H}$ with $\|u\|_2=1$,
\begin{align}\nonumber
E_a(u)&\ge \inte  V(x)|u(x)|^2 dx+\frac{a^*-a}{2}\inte |u(x)|^4 dx\\ \nonumber
&= \gamma + \inte \left[\left(V(x)-\gamma \right)|u(x)|^2 + \frac{a^*-a}{2}|u(x)|^4\right]dx\\
&\ge \gamma - \frac{1}{2(a^*-a)}\inte \left[\gamma -V(x)\right]^2_+dx\,,  \label{ib1}
\end{align}
where $[\,\cdot\,]_+ = \max\{0,\,\cdot\,\}$ denotes the positive part. For small enough $\gamma$, the set
$$
\{ x\in \R^2 \, : \, V(x) \leq \gamma \}
$$
is contained in the disjoint union of $n$ balls of radius at most $K \gamma^{1/p}$, centered at the minima $x_i$, for a suitable constant $K>0$. Moreover, $V(x) \geq (|x-x_i|/K)^p$ on these balls.  Hence
$$
\inte \left[\gamma -V(x)\right]^2_+ dx \leq n \int_{\R^2} \left[\gamma - (|x|/K)^p \right]_+^2dx = C \gamma^{2+2/p}
$$
with $C = n K^2 \pi p^2/[(p+1)(p+2)]$.  The lower bound in (\ref{4:con:1}) therefore follows from (\ref{ib1}) by taking $\gamma$ to be equal to  $[p(a^*-a)/(C(1+p))]^{p/(2+p)}$.

Next we shall prove the upper bound in (\ref{4:con:1}). We proceed similarly to the proof of Theorem~\ref{thm1}, and use a trial function of the form (\ref{def:trial}), with $0\leq \varphi\leq 1$. Recall the definition of $\mathcal{Z}$ in (\ref{def:Z}), and pick  $x_0 \in \mathcal{Z}$. We choose $R$ small enough so that
$$
V(x) \leq C |x-x_0|^p \quad \text{for $|x-x_0|\leq R$,}
$$
in which case we have
$$
\int_{\R^2} V(x) u(x)^2 dx \leq C \tau^{-p} A_{R,\tau}^2 \int_{\R^2} |x|^p Q(x)^2 dx \,.
$$
From the estimates (\ref{up1})--(\ref{W:est1})
we thus conclude that, for large $\tau$,
$$
e(a) \leq  \frac{\tau^2}{2 (a^*)^2} \left( a^*-a \right) \inte Q(x)^4dx + C \tau^{-p}  \int_{\R^2} |x|^p Q(x)^2 dx + O( \tau^{-\infty})\,.
$$
By taking $\tau =(a^*-a)^{-\frac{1}{p+2}}$, we arrive at the desired upper bound.
This completes the proof of the lemma.
\qed
\bigskip

Let $u_a$ be a  minimizer of (\ref{f}).
The following bound on the  $L^4(\R^2)$ norm of $u_a$ is a simple consequence of Lemma \ref{4:lem1}.

\begin{lem}\label{4:lem2}
Suppose  $u_a $ is  a  minimizer of (\ref{f}) with $V$ satisfying (\ref{as:v}). Then  there exists a positive constant  $K$, independent of $a$, such that
\begin{equation}
0<K(a^*-a)^{-\frac{2}{p+2}}\le \inte |u_a(x)|^4dx \le \frac{1}{K}(a^*-a)^{-\frac{2}{p+2}}\ \ \mbox{for}\ \ 0\leq a<  a^*.
\label{4:con:2'}
\end{equation}
\end{lem}

\noindent{\bf Proof.} Since by (\ref{key:min}) and (\ref{GNineq})
\[
 e(a)\ge \frac{a^*-a}{2}\inte |u_a(x)|^4dx\,,
\]
the upper bound in (\ref{4:con:2'}) follows immediately from Lemma \ref{4:lem1}.

To prove the lower bound in (\ref{4:con:2'}), we pick a $0<b<a$ and use that
\[
e(b)\le E_b(u_{a}) =  e(a)+ \frac{a-b}2 \int_{\R^2} |u_{a}(x)|^4 dx \,.
\]
By applying  Lemma \ref{4:lem1}, the above inequality implies that there exist two positive constants $m<M$ such that for any $0<b<a<a^*$,
\begin{equation}
\frac{1}{2}\inte |u_{a}(x)|^4dx\ge \frac{e(b)- e(a)}{a-b}\ge \frac{m(a^*-b)^{p/(2+p)} -M (a^*-a)^{p/(2+p)} }{a-b}\,.
\label{4:4:1}
\end{equation}
With $b = a -\gamma(a^*-a)$,  we can write the right side of (\ref{4:4:1}) as
$$
(a^*-a)^{-2/(2+p)} \frac{m (1+\gamma)^{p/(2+p)} - M}{\gamma}\,.
$$
The last fraction is positive for $\gamma$ large enough. For $a$ close to $a^*$, this then gives the desired lower bound. For smaller $a$, one can simply use the fact that  $\int |u_a(x)|^4 dx \geq \int |u_0(x)|^4 dx$ for any $0\leq a\leq a^*$, which follows from concavity of $e(a)$ (or from the bounds $e(a)\le E_a(u_0)$ and $e(0)\le E_0(u_a)$). This completes the proof of the lemma.
\qed
\bigskip

Let $u_a$ be a non-negative minimizer of (\ref{def:ea}), and define
\begin{equation}\label{def:eps}
 \eps :=(a^*-a)^{\frac{1}{2+p }}>0 \,.
\end{equation}
From (\ref{GNineq}) we conclude that
\[
e(a)\ge \Big(1-\frac{a}{a^*}\Big)\inte |\nabla u_a(x)|^2dx+\inte V(x)u_a(x)^2dx\,,
\]
and hence it follows from Lemma \ref{4:lem1} that
\begin{equation}
 \inte |\nabla u_a(x)|^2dx \le C\eps ^{-2}\quad \mbox{and}\quad \inte V(x)u_a(x)^2dx \le C\eps ^p.
\label{4:new:1}
\end{equation}

 For $1\leq
i\leq n$, define the $L^2(\R^2)$-normalized functions
\begin{equation}
w_a^{(i)}(x):=\eps u_a\left(\eps x+x_i\right)\,.
\label{4:4:2b}
\end{equation}
From (\ref{4:new:1}) and Lemma~\ref{4:lem2}, we have
\begin{equation}
 0<K \le \inte w_{a}^{(i)}(x)^4dx \le \frac{1}{K} \ , \quad \inte |\nabla w_{a}^{(i)}(x)|^2 dx \leq C
\label{4:thm:1}
\end{equation}
and also
\begin{equation}
\int _{\R ^2} V(x_i + \varepsilon x)w_{a}^{(i)}(x)^2 dx \leq C \eps^p \,.
\label{4:thm:2}
\end{equation}
In particular, the functions $w_a^{(i)}$ are bounded uniformly in $H^1(\R^2)$.

For any $\gamma >0$, we have
\[
\int_{\{V(x)\ge \gamma \eps ^p\}}u_{a}(x)^2 dx\le \frac{1}{\gamma \eps ^p}\inte V(x)u_{a}(x)^2 dx\le \frac{C}{\gamma}\,.
\]
For $\eps$ small enough, i.e., for $a$ sufficiently close to $a^*$, the set $\{x\in \R^2:\, V(x)\le\gamma \eps ^p \}$ is contained in $n$ disjoint balls with radius at most $C\gamma ^{1/p}\eps$, for some $C>0$, centered  at the points $x_i$. We thus deduce from the above inequality that
\begin{align*}
\frac{C}{\gamma}&\ge \int_{\{V(x)\ge \gamma \eps ^p\}}u_a(x)^2 dx=1- \int_{\{V(x)\le \gamma \eps ^p\}}u_a(x)^2 dx\\
&\ge  1- \sum^n_{i=1}\int _{\{|x-x_i|\le C\gamma ^{1/p}\eps\}}u_a(x)^2dx=1-
\sum ^n_{i=1}\int _{\{|x|\le C\gamma ^{1/p}\}} w^{(i)}_a(x)^2dx \,.
\end{align*}
In particular,
\begin{equation}\label{fc}
1\ge \sum ^n_{i=1}\int _{\{|x|\le C\gamma ^{1/p}\}} w^{(i)}_{a}(x) ^2dx\ge 1-\frac{C}{\gamma}\,.
\end{equation}

Since the functions $w_a^{(i)}$ are uniformly bounded in $H^1(\R^2)$, we can pass to a subsequence such that
\begin{equation}\label{wl}
w_{a}^{(i)} \rightharpoonup w_0^{(i)}  \ \text{weakly in $H^1(\R^2)$} \ , \quad 1\leq i \leq n\,,
\end{equation}
for suitable functions $w_0^{(i)}\in H^1(\R^2)$.
By Lemma~\ref{2:lem1}, the convergence holds strongly in $L^q(\{|x|\le C\gamma ^{1/p}\})$ for any $2\leq q<\infty$.
In particular, from (\ref{fc}) we conclude that
\[
1\ge \sum ^n_{i=1}\int _{\{|x|\le C\gamma ^{1/p}\}} w^{(i)}_0(x)^2dx\ge 1-\frac{C}{\gamma}\,.
\]
Since this bound holds for any $\gamma >0$, we finally conclude that
\begin{equation}
\sum ^n_{i=1}\|w^{(i)}_0\|_2^2=1 \,.\label{4:sum}
\end{equation}

Since $u_a$ is a minimizer of (\ref{def:ea}), it satisfies the Euler-Lagrange equation
\begin{equation}
-\Delta u_a(x) +V(x)u_a(x) =\mu_a u_a(x) +au_a(x) ^3 \quad \text{in $\R^2$}
\label{4:4:2a}
\end{equation}
for $\mu _a\in \R$ a suitable Lagrange multiplier. In fact,
$$
\mu_a = e(a) - \frac a 2 \inte u_a(x)^4 dx \,.
$$
The functions $w_a^{(i)}$ in (\ref{4:4:2b}) are thus non-negative solutions of
\begin{equation}
-\Delta w_a^{(i)}(x) +\eps^{2}V(x_i + \eps x)w_a^{(i)}(x) =\eps^{2}\mu_a w_a^{(i)}(x) +a w_a^{(i)}(x)^3 \quad \mbox{in}\quad \R^2\,.
\label{4:4:2}
\end{equation}
It follows from Lemma~\ref{4:lem2} that $\eps^2 \mu_a$ is uniformly bounded as $a \to a^*$, and strictly negative for $a$ close to $a^*$. By passing to a subsequence, if necessary, we can thus assume that $\eps^2\mu_a$ converges to some number $-\beta^2 < 0$ as $a\to a^*$. By passing to the weak limit (\ref{wl}), we see that the non-negative functions $w_0^{(i)}$ satisfy
\begin{equation}
-\Delta w^{(i)}_0(x)= -\beta^2  w^{(i)}_0(x) +a^* w^{(i)}_0(x)^3\,.
\label{4:4:5}
\end{equation}
By the maximum principle, either $w_0^{(i)}= 0$ identically, or
otherwise $w_0^{(i)}>0$ for all $x\in \R^2$. In the latter case, a
simple rescaling together with the uniqueness of positive solutions of
(\ref{Kwong}) up to translations allows us to conclude that
\begin{equation}\label{for}
w^{(i)}_0(x) =  \frac{\beta }{\|Q\|_2}Q(\beta (x-y_i)) \quad \text{for some $y_i\in \R^2$\,.}
\end{equation}
In particular, either $w_0^{(i)}=0$ or $\|w^{(i)}_0\|^2_2=1$. Because of (\ref{4:sum}), we see that exactly {\em one} $w_0^{(i)}$ is of the form (\ref{for}), while all the others are zero.

Let $1\leq j\leq n$ be such that $\|w_0^{(j)}\|_2=1$. From the norm
preservation we conclude that $w_a^{(j)}$ converges to $w_0^{(j)}$
{\em strongly} in $L^2(\R^2)$ and, in fact, strongly in $L^q(\R^2)$
for any $2\leq q<\infty$ because of $H^1(\R^2)$ boundedness. By going
to a subsequence, if necessary, we can also assume that the
convergence holds pointwise almost everywhere.

To complete the proof of Theorem \ref{thm3}, we compute
\begin{align*}
e(a) = E_a(u_a) & = \frac 1 {\eps^2} \left[  \inte |\nabla w_a^{(j)}(x)|^2 dx - \frac {a^*} 2 \inte w_a^{(j)}(x)^4 dx \right] \\ &  \quad + \frac{\eps^p}2 \inte w_a^{(j)}(x)^4 dx + \inte V(x_j+\eps x) w_a^{(j)}(x)^2 dx\,.
\end{align*}
The term in square brackets is non-negative and can be dropped for a lower bound. The $L^4(\R^2)$ norm of $w_a^{(j)}$ converges to the one of $w_0^{(j)}$, and from Fatou's Lemma it follows that
$$
\liminf_{\eps \to 0} \eps^{-p} \inte V(x_j+\eps x) w_a^{(j)}(x)^2 dx \geq  \kappa_j \inte |x|^{p} w_0^{(j)}(x)^2 dx\,,
$$
where $\kappa_j = \lim_{x\to x_j} V(x) |x-x_j|^{-p} \in (0,\infty]$. Moreover,
\begin{equation}\label{str}
\inte |x|^{p} w_0^{(j)}(x)^2 dx = \frac {1}{\beta^{p}\|Q\|_2^2}  \inte |x+y_j|^p Q(x)^2 dx \geq \frac {1}{\beta^{p}\|Q\|_2^2}  \inte |x|^p Q(x)^2 dx
\end{equation}
since $Q$ is a radial decreasing function.
In particular,
\begin{equation}\label{mb}
\liminf_{a\to a^*} \frac{e(a)}{(a^*-a)^{p/(2+p)}} \geq \frac 12 \|w_0\|_4^4 + \kappa_j \inte |x|^p w_0^{(j)}(x)^2 dx \geq
\frac{ 2}{ a^*} \left( \frac{\beta^2 }{2 } + \lambda_j^{2+p} \frac 1 {p \beta^p }\right)\,,
\end{equation}
where $\lambda_j$ is defined in (\ref{def:li}), and we have used that
$\|Q\|_4^4 = 2 \|Q\|_2^2 = 2 a^*$ (which follows from the fact that
$Q$ satisfies Eq.~(\ref{Kwong}) and equality in (\ref{GNineq})). Taking
the infimum over $\beta>0$ (which is achieved for $\beta = \lambda_j$)
yields
\begin{equation}\label{lim}
\liminf_{a\to a^*} \frac{e(a)}{(a^*-a)^{p/(2+p)}} \geq \frac{ \lambda^2}{a^*} \frac{p+2}{p}\,,
\end{equation}
where $\lambda = \min_j \lambda_j$, as before.

The limit in (\ref{lim}) actually exists, and is equal to the right side. To see this, one simply takes
$$
u(x) = \frac \beta{\eps \|Q\|_2} Q\left( \beta \frac{x-x_j}{\eps }\right)
$$
as a trial function for $E_a$, and minimizes over $1\leq j\leq n$ and $\beta>0$. The result is that
\begin{equation}\label{lime}
\lim_{a\to a^*} \frac{e(a)}{(a^*-a)^{p/(2+p)}} = \frac{\lambda^2}{a^*} \frac{p+2}{p}\,.
\end{equation}

From the equality (\ref{lime}) we can draw several conclusions. First,
the $j$ defined above is such that $\lambda_j=\lambda$, i.e.,
$x_j\in\mathcal{Z}$. Second, $\beta$ is unique (i.e., independent of
the choice of the subsequence) and equal to the expression minimizing
(\ref{mb}), i.e., $\beta = \lambda$. Finally, $y_j = 0$, since the
inequality (\ref{str}) is strict for $y_j\neq 0$. We have thus shown
that
$$
w_a^{(j)}(x) = \eps u_a(\eps x+ x_j)  \to \frac{\lambda}{\|Q\|_2} Q(\lambda x) \quad \text{as $a\to a^*$,}
$$
with $x_j \in \mathcal{Z}$. This completes the proof of
Theorem~\ref{thm3}.
\qed

\end{document}